# Parallel Algorithm for Frequent Itemset Mining on Intel Many-core Systems[*]

*Mikhail Zymbler*

South Ural State University; Russian Federation; mzym@susu.ru

**Abstract.** Frequent itemset mining leads to the discovery of associations and correlations among items in large transactional databases. Apriori is a classical frequent itemset mining algorithm, which employs iterative passes over database combining with generation of candidate itemsets based on frequent itemsets found at the previous iteration, and pruning of clearly infrequent itemsets. The Dynamic Itemset Counting (DIC) algorithm is a variation of Apriori, which tries to reduce the number of passes made over a transactional database while keeping the number of itemsets counted in a pass relatively low. In this paper, we address the problem of accelerating DIC on the Intel Xeon Phi many-core system for the case when the transactional database fits in main memory. Intel Xeon Phi provides a large number of small compute cores with vector processing units. The paper presents a parallel implementation of DIC based on OpenMP technology and thread-level parallelism. We exploit the bit-based internal layout for transactions and itemsets. This technique reduces the memory space for storing the transactional database, simplifies the support count via logical bitwise operation, and allows for vectorization of such a step. Experimental evaluation on the platforms of the Intel Xeon CPU and the Intel Xeon Phi coprocessor with large synthetic and real databases showed good performance and scalability of the proposed algorithm.

## 1. Introduction

*Association rule mining* is one of the important problems in data mining [1]. The task is to discover strong associations among the items from a transactional database such that the occurrence of one item in a transaction implies the occurrence of another. Association rule mining is divided into two subtasks [1]. The first one is to find all *frequent itemsets* that consist of items, which often occur together in transactions. The second one is to generate all the *association rules* from the frequent itemsets found.

---

[*] This article is an extended version of the paper [23], which was recommended for the publication by the MiproBIS'2017 Program Committee as the best paper of the conference.

In this paper, we address the task of frequent itemset mining, which can be formally described as follows. Let $\mathcal{I} = (i_1, \ldots, i_m)$ be a set of literals, called *items*. Let $\mathcal{D} = (T_1, \ldots, T_n)$ be a *database of transactions*, where each *transaction* $T_i \subseteq \mathcal{I}$ consists of a set of items (*itemset*). An itemset that contains $k$ items is called a $k$-itemset. The *support* of an itemset $I \subseteq \mathcal{I}$ denotes the fraction of transactions in $\mathcal{D}$ that contain the itemset $I$. If the support of an itemset $I \subseteq \mathcal{I}$ satisfies the user-specified minimum support threshold called $minsup$, then $I$ is *frequent itemset*. Let the set of all frequent $k$-itemsets be denoted by $\mathcal{L}_k$ and $\mathcal{L} = \bigcup_{k=1}^{k=k_{max}} \mathcal{L}_k$ denotes a set of all frequent itemsets, where $k_{max}$ is the number of items in the longest frequent itemset. Given the transactional database $\mathcal{D}$ and minimum support threshold $minsup$, the goal of frequent itemset mining is to find the set of all frequent itemsets $\mathcal{L}$.

There is a wide spectrum of algorithms for frequent itemset mining, and none of them outperforms all others for all possible transactional databases and values of $minsup$ threshold [9]. *Apriori* [1] is one of the most popular itemset mining algorithms, for which many refinements and parallel implementations for various platforms were proposed. *Dynamic Itemset Counting (DIC)* [3] is a variation of Apriori, which tries to reduce the number of passes made over a transactional database while keeping the number of itemsets counted in a pass relatively low. Despite the fact that DIC has good potential of parallelization [3], it still has not been implemented for modern Intel many-core systems, to the best of our knowledge.

In this paper, we address the problem of accelerating the DIC algorithm on the Intel Xeon Phi many-core system. *Intel Xeon Phi* [21] provides a large number of small compute cores with a high local memory bandwidth. Each core supports a computational power weaker than that of the Intel Xeon core and provides 512-bit wide vector processing unit (VPU). VPU supports data-level parallelism by a set of vector instructions, thanks to which it is possible to load and calculate several numbers at once (e.g. eight 64-bit integers or sixteen 16-bit floats). Such a routine is called *vectorization*, and Intel compilers provide options for automatic vectorization. Since Intel Xeon Phi is based on Intel x86 architecture, it supports the same programming tools as a regular Intel Xeon CPU. Thus, Intel Xeon Phi can be considered as an attractive hardware platform for the thread-level parallel algorithm.

The basic contribution of the paper is as follows. We propose a parallel implementation of the DIC algorithm for the Intel Xeon Phi many-core system. We exploit a bit-based internal layout for transactions and itemsets assuming that such a representation of a transactional database fits in main memory. This technique reduces memory space of storing the transactional database and simplifies the support count and generation of potentially frequent candidate itemsets via logical bitwise operations. The algorithm is parallelized using OpenMP technology and thread-level parallelism. We conduct experiments on large synthetic and real databases to evaluate the performance and scalability of our algorithm.

The rest of the paper is organized as follows. In section 2, related work is discussed. Section 3 provides a brief description of the original DIC algorithm. The proposed parallel algorithm is presented in Section 4. The results of experimental evaluation of the algorithm are described in Section 5. Finally, Section 6 concludes the paper.

## 2. Related work

The original DIC algorithm was presented by Brin *et al.* in [3], where the authors briefly discuss a way to parallelize DIC using the distribution of the transactional database among the nodes so that each node counts all itemsets for its own data segment.

Paranjape-Voditel *et al.* proposed DIC-OPT [14], a parallel version of DIC for distributed memory systems. The key idea is that each node sends messages with the counts of potentially frequent itemsets to other nodes after every block of M transactions has been read. This initiates the early counting of the itemsets on other nodes without waiting for synchronization with other nodes. The authors carried out experiments on up to 12 nodes where their implementation showed sublinear speedup.

Cheung *et al.* suggested APM [5], a DIC-based parallel algorithm for SMP systems. APM is an adaptive parallel mining algorithm, where all CPUs generate candidates dynamically and count itemset supports independently without synchronization. The transactional database is partitioned across CPUs with highly homogeneous itemset distributions. This technique addresses to the problem of a large number of candidates because of the low homogeneous itemset distribution in most cases. The experiments on the Sun Enterprise 4000 server with up to 12 nodes showed that APM outperforms Apriori-like parallel algorithms. However, the APM speedup gradually drops down to 4 when the number of nodes is greater than four. This is because APM suffers from the SMP system inherent problem of I/O contention when the number of nodes is large.

Schlegel *et al.* proposed mcEclat [19], a parallel version of Eclat [22] for the Intel Xeon Phi coprocessor. mcEclat converts a dataset being mined into a set of tid-bitmaps, which are repeatedly intersected to obtain the frequent itemsets. Tid-bitmap maps the IDs of transactions, in which an itemset exists, to bits in a bitmap at certain positions. Tid-bitmaps are intersected via logical bitwise AND operation and then the support of an itemset is obtained by counting the bits set to one in its respective tid-bitmap. Experiments showed up to 100 times speedup of mcEclat on Intel Xeon Phi. However, the algorithm performance on the Intel Xeon Phi coprocessor is similar or slightly worse (for smaller values of $minsup$) than on the system with two Intel Xeon CPUs when the maximum number of threads is employed on both systems. The reason is that mcEclat does not fully exploit the vector processing capabilities of Intel Xeon Phi.

Kumar *et al.* presented Bitwise DIC [12], a serial version of the DIC algorithm based upon tid-bitmap technique mentioned above. Authors report that Bitwise DIC outruns the original DIC on datasets with up to 5,000 transactions for the fixed $minsup$ value.

In serial algorithms, MAFIA [4] and BitTableFI [7], Burdick *et al.* and Dong *et al.*, respectively, employed vertical bitmap to compress the transactional database for quick candidate itemsets generation and the support count. Vertical bitmap is a set of integers in which every bit represents an item. If an item $i$ appears in the $j$-th transaction, then the $j$-th bit of the bitmap for the item $i$ is set to one; otherwise, the bit is set to zero. This idea is applied to both transactions and itemsets. In the case when itemsets appear in a significant number of transactions, the vertical bitmap is the smallest representation of the information. However, the weakness of a vertical representation is the sparseness of the bitmaps, especially at the lower support levels.

## 3. Serial DIC Algorithm

*Dynamic Itemset Counting (DIC)* [3] is a variation of the most well-known *Apriori algorithm* [1]. Apriori is an iterative, level-wise algorithm, which uses a bottom-up search. At the first pass over transactional database, it processes 1-itemsets and finds $\mathcal{L}_1$ set. A subsequent pass $k$ consists of two steps, namely candidate generation and pruning. At the *candidate generation* step, Apriori combines elements of $\mathcal{L}_{k-1}$ set to form potentially frequent *candidate* $k$-itemsets. At the *pruning* step, it discards infrequent candidates using the *a priori* principle, which states that any infrequent ($k$-1)-itemset cannot be a subset of a frequent $k$- itemset. Apriori counts support of candidates, which have not been pruned, and proceeds with such passes until there are no more candidates after pruning.

The DIC algorithm tries to reduce the number of passes made over the transactional database while keeping the number of itemsets counted in a pass relatively low. Algorithm 1 depicts pseudo-code of DIC. The algorithm processes the database with stops at equal-length intervals between transactions specified by the $M$ parameter of the algorithm. At the end of the transactional database, it is necessary to rewind to its beginning.

---

**Algorithm** SERIALDIC
**Input:** $\mathcal{D}$, $minsup$, $M$
**Output:** $\mathcal{L}$
           ▷ Initialize sets of itemsets
SOLIDBOX←∅; SOLIDCIRCLE←∅; DASHEDBOX←∅
DASHEDCIRCLE ← $\mathcal{I}$
**while** DASHEDCIRCLE ∪ DASHEDBOX ≠ ∅ **do**
      ▷ Scan database and rewind if necessary
 Read($\mathcal{D}$, $M$, $Chunk$)
 **if** EOF($\mathcal{D}$) **then**
  Rewind($\mathcal{D}$)
 **for all** $T \in Chunk$ **do**
          ▷ Count support of itemsets
  **for all** $I \in$ DASHEDCIRCLE ∪ DASHEDBOX **do**
   **if** $I \subseteq T$ **then**
    $support(I) \leftarrow support(I) + 1$
         ▷ Generate candidate itemsets
  **for all** $I \in$ DASHEDCIRCLE **do**
   **if** $support(I) \geq minsup$ **then**
    MoveItemset($I$, DASHEDBOX)
   **for all** $i \in \mathcal{I}$ **do**
    $C \leftarrow I \cup i$
    **if** $\forall s \subseteq C$ $s \in$ SOLIDBOX ∪ DASHEDBOX **then**
     MoveItemset($I$, DASHEDCIRCLE)
         ▷ Check full pass completion
  **for all** $I \in$ DASHEDCIRCLE ∪ DASHEDBOX **do**
   **if** IsPassCompleted($I$) **then**
    **switch** Shape($I$)
     Dashed: MoveItemset($I$, DASHEDBOX)
     Solid: MoveItemset($I$, SOLIDBOX)
$\mathcal{L} \leftarrow$ SOLIDBOX

*Algorithm 1. Serial DIC algorithm.*

DIC maintains four sets of itemsets, namely *Dashed Circle*, *Dashed Box*, *Solid Circle* and *Solid Box*. Itemsets in the "dashed" sets are subjects for the support count while itemsets in the "solid" sets do not need to be counted. "Circles" contain infrequent itemsets while "boxes" contain frequent itemsets. Thus, *Dashed Circle* and *Dashed Box* contain itemsets that are suspected infrequent and suspected frequent, respectively, while *Solid Circle* and *Solid Box* contain itemsets that are confirmed infrequent and confirmed frequent, respectively. Figure 1 depicts lifecycle of an itemset in the DIC algorithm.

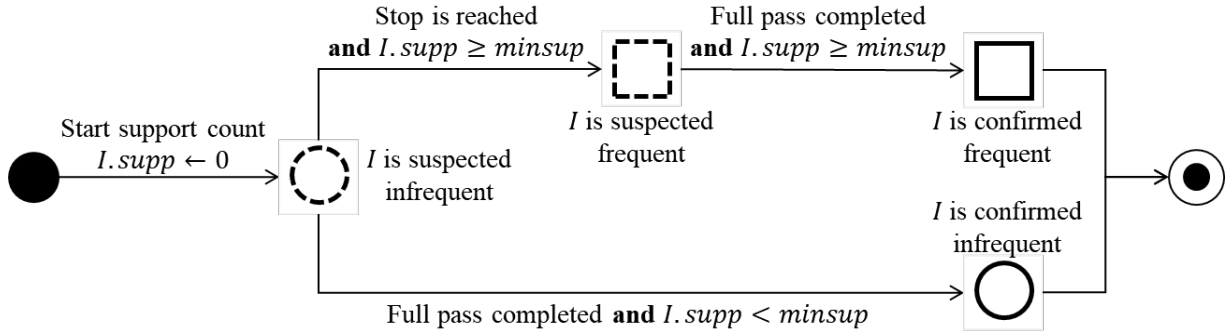

*Figure 1*. Lifecycle of an itemset in the DIC algorithm.

At start, *Dashed Box*, *Solid Circle*, and *Solid Box* are assumed empty, and *Dashed Circle* contains all 1-itemsets. Before the stop, DIC counts support of itemsets from "dashed" sets for each transaction. At any stop, DIC performs as follows. Itemsets whose support exceeds *minsup* are moved from *Dashed Circle* to *Dashed Box*. New itemsets are added into *Dashed Circle*, they are immediate supersets of those itemsets from *Dashed Box* with all of its subsets from "box" sets. Itemsets that have completed one full pass over the transactional database are moved from the "dashed" set to "solid" set. DIC proceeds as long as itemsets remain in the "dashed" sets.

## 4. Parallel DIC Algorithm

### 4.1. Internal Data Layout

In this paper, we propose the direct bit representation for both transactions and itemsets. This means that the transaction $T \subseteq \mathcal{D}$ (an itemset $I \subseteq \mathcal{I}$, respectively) is represented by a word where each $p$-th bit is set to one if an item $i_p \in T$ ($i_p \in I$, respectively) and all other bits are set to zero. The word length $W$ in bytes depends on the system environment and is calculated as $W = \left\lceil \frac{m}{sizeof(\text{byte})} \right\rceil$. In our implementation, we use C++ and unsigned long long int data type, so we have $W=8$ and $m=64$.

Let us denote by $BitMask$ a function that returns direct bit representation of a given itemset or transaction as a word, i.e. $BitMask: \mathcal{I} \to \mathbb{Z}_+$. Then, the direct bit representation of transactional database $\mathcal{D}$ is an $n$-element array $\mathcal{B}$ where $\mathcal{B}_j = BitMask(T_j) \ \forall j \in 1..n$.

The direct bit representation has several major merits. It often requires less space than byte-based representation for dense transactional database with long transactions. In fact, $\mathcal{B}$ requires $n \cdot W$ bytes to store and allows $\mathcal{B}$ to fit in main memory. For instance, *netflix*, one of the most referenced datasets, contains $n$=17,771 transactions consisting of $m$=480,189 distinct items. Hence, the direct bit representation of the *netflix* dataset takes about 1 Gb. Thus, we further assume that $\mathcal{B}$ is preliminary produced from $\mathcal{D}$ and available in main memory.

Additionally, the direct bit representation simplifies support counting and vectorization of this operation. The fact that an itemset $I$ exists in a transaction $T$ (i.e. $I \subseteq T$) can be checked by one logical bitwise operation, that is $BitMask(I)$ AND $BitMask(T) = BitMask(I)$. Such an implementation allows for auto-vectorization of the support count loop by the compiler.

Thereby, we implement an itemset as a record structure with the following basic fields, namely $mask$ to provide direct bit representation, $k$ as number of items in the itemset, $stop$ as counter to determine when full pass for the given itemset is completed, and $supp$ to store support count.

To implement a set of itemsets, we use vector, which represents an array of elements belonging to the same type and provides random access to its elements with an ability to automatically resize when appending elements. Such a data structure is implemented in C++ Standard Template Library as a class with iterator and methods for inserting an element and removing an element with complexity of $O(1)$ and $O(s)$ respectively, where $s$ is the current size of a vector.

In order to reduce costs of moving elements across vectors, we establish a $DASHED$ vector for "dashed box" and "dashed circle" itemsets and a $SOLID$ vector for "solid box" and "solid circle" itemsets, and provide the itemset record structure with the $shape$ field to indicate an appropriate set the given itemset belongs to.

### 4.2. Parallelization of the Algorithm

The proposed parallel version of DIC algorithm (hereinafter ParallelDIC) is presented in Algorithm 2, and basic sub-algorithms are depicted in Algorithm 3, Algorithm 4, and Algorithm 5.

We enhance the classical DIC algorithm by adding two more stages, namely *FirstPass* and *Prune* where each of them is aimed to reducing the number of itemsets to perform support counting.

We parallelize the following stages of the algorithm, namely the support count (cf. Algorithm 3), pruning of the *Dashed Circle* set (cf. Algorithm 4) and check of full pass completion for itemsets (cf. Algorithm 5) through OpenMP technology and thread-level parallelism.

In the classical DIC (cf. Algorithm 1), the *Dashed Circle* set is initialized by all 1-itemsets. In contrast, we use the technique of full first pass [5]. This means that we initially perform one full pass over $\mathcal{D}$ to find $\mathcal{L}_1$, the set of frequent 1-itemsets (this done similarly to Algorithm 3). Then candidate 2-itemsets are computed from $\mathcal{L}_1$ through the Apriori join procedure [1]. This done via logical bitwise OR operation on each pair of frequent 1-itemsets, and candidates are inserted in the *Dashed Circle* set. This technique helps to reduce cardinality of the *Dashed Circle* set in further computations because infrequent 1-itemsets and their supersets have been pruned according to the *a priori* principle.

**Algorithm** PARALLELDIC  
**Input:** $\mathcal{B}, minsup, M, num\_of\_threads$  
**Output:** $\mathcal{L}$  
$\triangleright$ Initialize sets of itemsets  
$SOLID.init(); DASHED.init()$  
**for all** $i \in 0..m-1$ **do**  
  $I.shape \leftarrow \text{NIL}; \text{SETBIT}(I.mask, i)$  
  $I.stop \leftarrow 0; I.supp \leftarrow 0; I.k \leftarrow 1$  
  $SOLID.push\_back(I)$  
$k \leftarrow 1; stop \leftarrow 0; stop_{max} \leftarrow \left\lceil \frac{n}{M} \right\rceil$  
FIRSTPASS($SOLID, DASHED$)  
**while not** $DASHED.empty()$ **do**  
  $\triangleright$ Scan database and rewind if necessary  
  $stop \leftarrow stop + 1$  
  **if** $stop > stop_{max}$ **then**  
    $stop \leftarrow 1$  
  $first \leftarrow (stop-1) \cdot M; last \leftarrow stop \cdot M - 1$  
  $k \leftarrow k + 1$  
  COUNTSUPPORT($DASHED, num\_of\_threads$)  
  PRUNE($DASHED, minsup$)  
  MAKECANDIDATES($DASHED$)  
  CHECKFULLPASS($DASHED, minsup$)  
$\mathcal{L} \leftarrow \{I \mid I \in SOLID \land I.shape = \text{BOX}\}$

*Algorithm 2.* Parallel DIC algorithm.

The original algorithm performs support counting by two nested loops where the outer loop takes transactions and the inner loop takes the "dashed" itemsets. As opposed to DIC, we change the order of these loops (cf. Figure 2). This shuffle allows avoiding data races when threads process different transactions and need to change the support count of the same itemsets simultaneously.

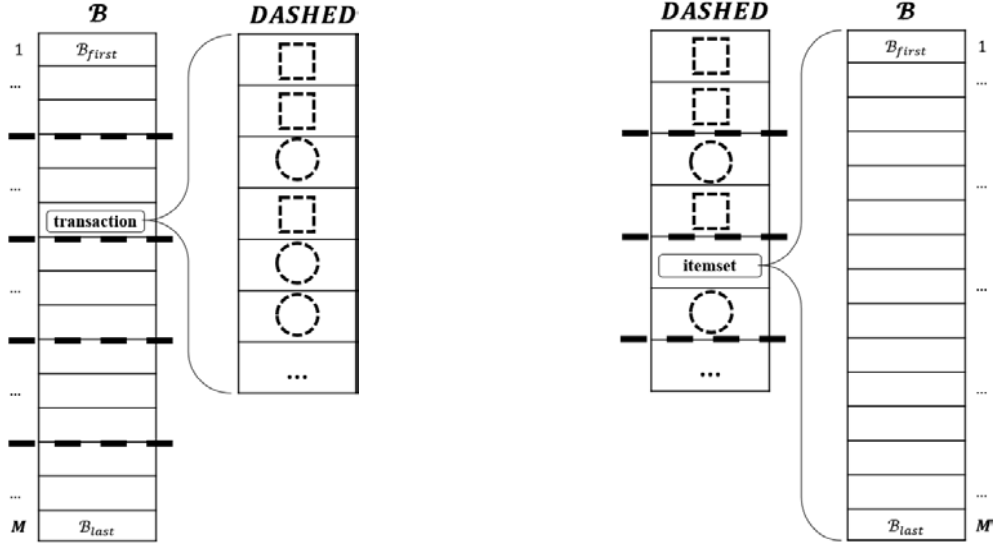

a) Count all the "dashed" itemsets for each transaction: data races among threads are possible

b) Count all the transactions for each "dashed" itemset: no data races among threads

*Figure 2.* Support count in the ParallelDIC algorithm.

Then, we parallelize the outer loop through omp parallel for pragma (cf. Algorithm 3).

---

**Algorithm** COUNTSUPPORT
**Input:** $DASHED, num\_of\_threads$
**Output:** $DASHED$
**if** $DASHED.size() \geq num\_of\_threads$ **then**
   #pragma omp parallel for
   **for all** $I \in DASHED$ **do**
     $I.stop \leftarrow I.stop + 1$
     **for all** $T \in \mathcal{B}_{first}..\mathcal{B}_{last}$ **do**
       **if** $I.mask$ **AND** $T = I.mask$ **then**
         $I.supp \leftarrow I.supp + 1$
**else**
   omp_set_nested(true)
   #pragma omp parallel for num_threads($DASHED.size()$)
     **for all** $I \in DASHED$ **do**
       $I.stop \leftarrow I.stop + 1$
       #pragma omp parallel for reduction(+: $I.supp$) num_threads($\left\lceil\frac{num\_of\_threads}{DASHED.size()}\right\rceil$)
       **for all** $T \in \mathcal{B}_{first}..\mathcal{B}_{last}$ **do**
         **if** $I.mask$ **AND** $T = I.mask$ **then**
           $I.supp \leftarrow I.supp + 1$

---

*Algorithm 3.* Support count sub-algorithm.

Additionally, our algorithm balances the load of threads depending on the current total number of elements in both *Dashed Circle* and *Dashed Box* sets (cf. Figure 3).

If the number of available threads does not exceed the current total number of "dashed" itemsets, we parallelize the outer loop (along itemsets) using all threads. Otherwise, we enable nested parallelism and parallelize the outer loop using a number of threads equal to the current total number of "dashed" itemsets. Then we parallelize the inner loop (along transactions) so that each outer thread forks an equal-sized set of descendant threads where descendants perform counting by reducing the summing operation. This balancing technique allows for processing data effectively in the final stage of counting when the number of candidate itemsets tends to zero and increases the overall performance of the algorithm.

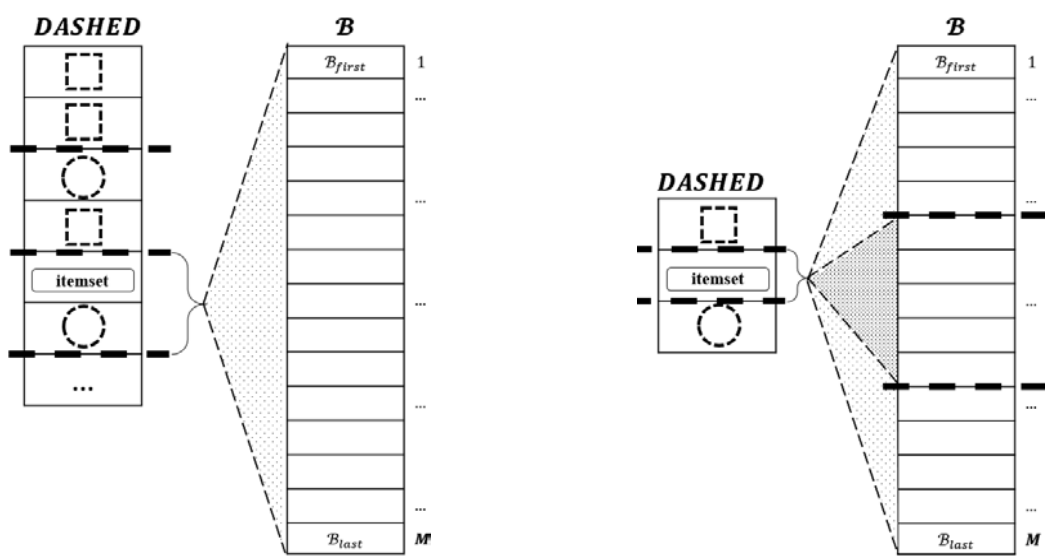

a) Number of threads is less than number of "dashed" itemsets: a thread takes its "dashed" itemsets to count

b) Number of threads is greater than number of "dashed" itemsets: a thread takes one "dashed" itemset and forks descendant threads to count

*Figure 3*. Load Balancing in the ParallelDIC algorithm.

After the support count, in addition to moving appropriate itemsets from *Dashed Circle* set to *Dashed Box* set as in classical DIC, we reduce the *Dashed Circle* set pruning clearly infrequent itemsets as follows [12]. We compute an itemset highest possible support by adding its current support to the number of transactions have not been processed yet (cf. Algorithm 4). If the value of the itemset highest possible support is less than *minsup* threshold, then the itemset is pruned, and after that, we prune all its supersets according to the *a priori* principle.

```
Algorithm PRUNE
Input: DASHED, num_of_threads
Output: DASHED
#pragma omp parallel for num_threads(num_of_threads)
for all I ∈ DASHED and I.shape = CIRCLE do
   if I.supp ≥ minsup then
      ▷ Move appropriate itemsets to Dashed Box set
      I.shape ← BOX
   else
      ▷ Prune clearly infrequent itemsets
      supp_max ← I.supp + M · (stop_max − I.stop)
      if supp_max < minsup then
         I.shape ← NIL
         for all J ∈ DASHED and J.shape = CIRCLE do
            if I.mask AND J.mask = I.mask then
               J.shape ← NIL
DASHED.erase({I|I.shape = NIL})
```

*Algorithm 4.* Pruning sub-algorithm.

After the reduction of the *Dashed Circle* set, we generate afresh itemsets to be inserted in that set performing Apriori join procedure [1] via the logical bitwise OR operation between all itemsets marked as "boxes".

```
Algorithm CHECKFULLPASS
Input: DASHED, num_of_threads
Output: DASHED
#pragma omp parallel for num_threads(num_of_threads)
for all I ∈ DASHED do
   if I.stop = stop_max then
      if I.supp ≥ minsup then
         I.shape ← BOX
      SOLID.push_back(I)
      I.shape ← NIL
DASHED.erase({I|I.shape = NIL})
```

*Algorithm 5.* Check full pass sub-algorithm.

Finally, for all itemsets in the *Dashed Circle* set, we check if an itemset has been counted through all transactions, and if yes, we make the itemset "solid" and stop counting it. If the itemset support equals to or exceeds the $minsup$ threshold, then we mark it as "box" (cf. Algorithm 5). This activity is also parallelized along itemsets through omp parallel for pragma.

In the end, $SOLID$ vector contains "box" itemsets as an output of the algorithm.

## 5. Experiments

### 5.1. Experimental Setup

*Measures.* In the experiments, we evaluated the speedup and parallel efficiency of the developed algorithm, where such characteristics of parallel-algorithm scalability are defined as follows. *Speedup* and *parallel efficiency* of a parallel algorithm employing $k$ threads are calculated, respectively, as $s(k) = \frac{t_1}{t_k}$ and $e(k) = \frac{s(k)}{k}$, where $t_1$ and $t_k$ are the run times of the algorithm when one and $k$ threads are employed, respectively.

*Competitors.* In previous work [23], our experiments showed that the performance of serial implementation of DIC in [8] substantially inferior to both our algorithm and serial Apriori in [2]. Thus, in this paper, we compared the performance of ParallelDIC with serial implementations of the following algorithms in [2]: Apriori, Eclat, and FP-Growth.

*Datasets.* Experiments in our previous work [23] also showed that, for datasets with hundreds of thousands of transactions (e.g. the SKIN [6] dataset and the RECORDLINK [18] dataset with 245,057 and 574,913 transactions, respectively), ParallelDIC demonstrates degradation of the speedup and parallel efficiency. This is because of the following reasons. For datasets with relatively small number of short transactions, our algorithm provides insufficient amount of work in support counting, which is the heaviest part of the algorithm. At the same time, efficiency of Intel Xeon Phi utilization as well as vectorization increase with the growth of the problem size [20]. Thus, in this paper, we evaluated our algorithm on two datasets, each of which contains tens of millions of transactions (cf. Table 1).

*Table 1*. Specifications of datasets.

| Dataset | Category | Transactions | | | Frequent itemsets ($minsup$=0.1) | |
|---|---|---|---|---|---|---|
| | | $n$ | $m$ | Avg. length | Total number | $k_{max}$ |
| 20M | Synthetic | $2 \cdot 10^7$ | 64 | 40 | 4,606 | 6 |
| Tornado20M | Real | $2 \cdot 10^7$ | 64 | 15 | 346 | 4 |

Synthetic dataset 20M was prepared through IBM Quest Data Generator [10] similar to the paper [3] where the original DIC algorithm was proposed. Eventually, the 20M dataset gives more than 4,600 frequent itemsets with at most 6 items.

Tornado20M is a real dataset with one-month voltage log of the Tornado SUSU supercomputer [11] nodes. Such a log is mined to discover the strong associations among the racks, shelves, and nodes of the supercomputer, and dangerous values of voltage. Tornado

SUSU consists of 8 racks, and each rack consists of 8 shelves, each with 6 nodes onboard. For each node, there are 4 possible values of measured voltage, and for each possible value there are 4 statuses (i.e. "less than norm", "norm", "greater than norm", and "error"). Thus, it is possible to code a transaction of such a log using 64 bits (i.e. 8 bits for the number of a rack, 8 bits for the number of a shelf, and 8 bits for each of 6 nodes where each pair of bits represents the status of the measured voltage). Eventually, the Tornado20M dataset gives more than 340 frequent itemsets with at most 4 items.

The experiments were carried out on the node of the Tornado SUSU supercomputer [11]. Such a node consists of a host, which is two 6-core Intel Xeon CPU, and a 61-core Intel Xeon Phi coprocessor. Table 2 depicts technical specifications of the hardware.

*Table 2*. Specifications of hardware.

| Specifications | Host | Coprocessor |
| --- | --- | --- |
| Model, Intel Xeon | X5680 | Phi SE10X |
| Number of physical cores | 2×6 | 61 |
| Hyper-threading factor | 2× | 4× |
| Number of logical cores | 24 | 244 |
| Frequency, GHz | 3.33 | 1.1 |
| Peak performance, TFLOPS | 0.371 | 1.076 |
| Memory, Gb | 24 | 8 |
| Cache, Mb | 12 | 30.5 |

*Parameters.* In the experiments, we took $M$, the number of transactions that should be processed before a stop, as $n/2$ in order to avoid overheads for initializing threads at each stop and increase the algorithm performance. We also evaluated the effect of the $minsup$ threshold on the algorithm speedup. As for the experiments studying the algorithm scalability, we took $minsup$ threshold as 0.1 as the most common value of support.

### 5.2. Results

Figure 4 illustrates the performance of ParallelDIC on both Intel Xeon and Intel Xeon Phi in comparison with serial Apriori, FP-Growth, and Eclat on Intel Xeon. Among serial implementations, Apriori performs the worst for the 20M dataset and performs the best for the Tornado20M dataset because this algorithm performance suffers when datasets with long transactions and the large number of frequent itemsets are processed, and may overtake competitors when transactions are relatively short and the number of frequent itemsets is small.

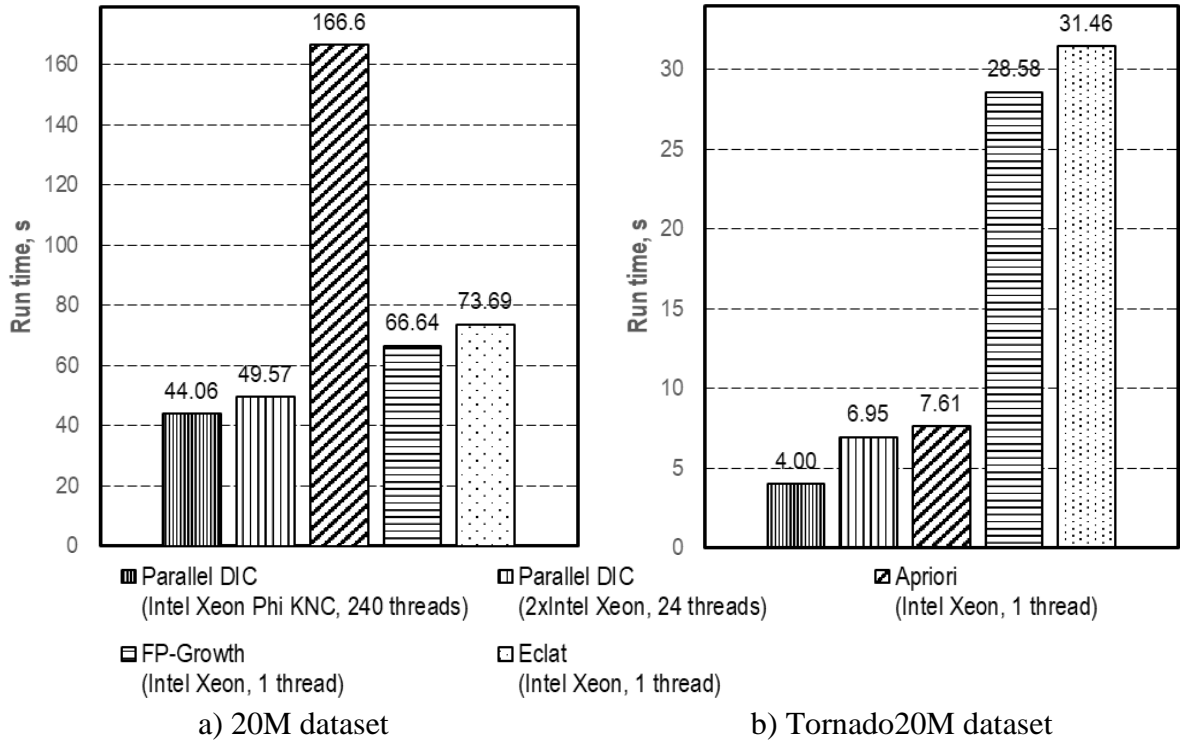

a) 20M dataset
b) Tornado20M dataset

*Figure 4.* Comparison of performance.

As we can see, ParallelDIC on Intel Xeon Phi outruns itself on two Intel Xeon up to 1.5 times. ParallelDIC on Intel Xeon Phi also outruns the best serial competitor on Intel Xeon up to 2 times. This is because more threads of Intel Xeon Phi allow for better exploiting the vectorization abilities of our algorithm.

In addition, we compare performance of ParallelDIC for the cases when the Intel compiler auto-vectorization option was enabled or disabled. Results in Table 3 show that, for the Tornado20M dataset, vectorization gives a performance boost of 1.2 and 2.6 times on Intel Xeon and Intel Xeon Phi, respectively.

*Table 3.* Performance of the algorithm depending on vectorization option of the compiler.

| Hardware | Run time, s when vectorization is | |
| --- | --- | --- |
| | *enabled* | *disabled* |
| Intel Xeon Phi | 4.00 | 10.36 |
| Intel Xeon | 6.95 | 8.55 |

Figure 5 depicts the speedup and parallel efficiency of ParallelDIC. On Intel Xeon Phi, our algorithm shows close-to-linear speedup and near 100% parallel efficiency, when the number of threads matches the number of physical cores the algorithm is running on. When the algorithm employs more than one thread per physical core, speedup becomes sublinear (it slows down to 88 and 108 for the 20M dataset and the Tornado20M dataset, respectively), and parallel efficiency diminishes accordingly (down to 37% and 45% with respect to a dataset). On two Intel Xeon, there is a similar tendency but with more moderate results for the Tornado20M dataset. For this dataset, the algorithm speedup and parallel efficiency drop to 8 and 35%, respectively, when the maximal possible number of threads per physical core is employed.

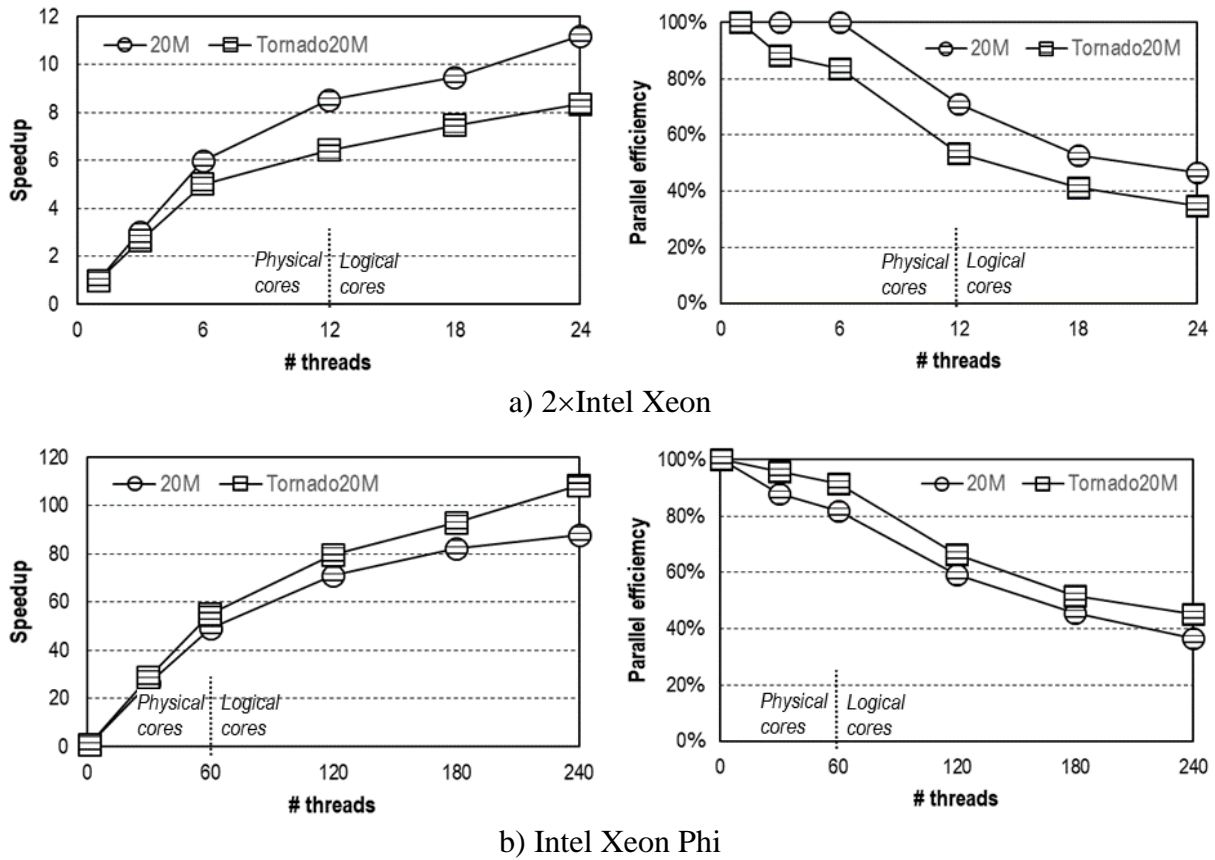

a) 2×Intel Xeon

b) Intel Xeon Phi

*Figure 5.* Speedup and parallel efficiency of the algorithm.

Figure 6 depicts speedup of the algorithm with respect to the $minsup$ threshold. As expected, on both platforms and for both datasets, the algorithm speedup suffers from decreasing of the $minsup$ value since this significantly increases the number of candidate itemsets to be counted. Our algorithm still shows better speedup when only physical cores are involved, and better speedup on Intel Xeon Phi system than on two Intel Xeon node.

Summing up, ParallelDIC demonstrates good performance and scalability for large datasets (about tens of millions of transactions) and for the most common value of minimum support threshold ($minsup$=0.1) on Intel many-core platforms, especially on the Intel Xeon Phi system.

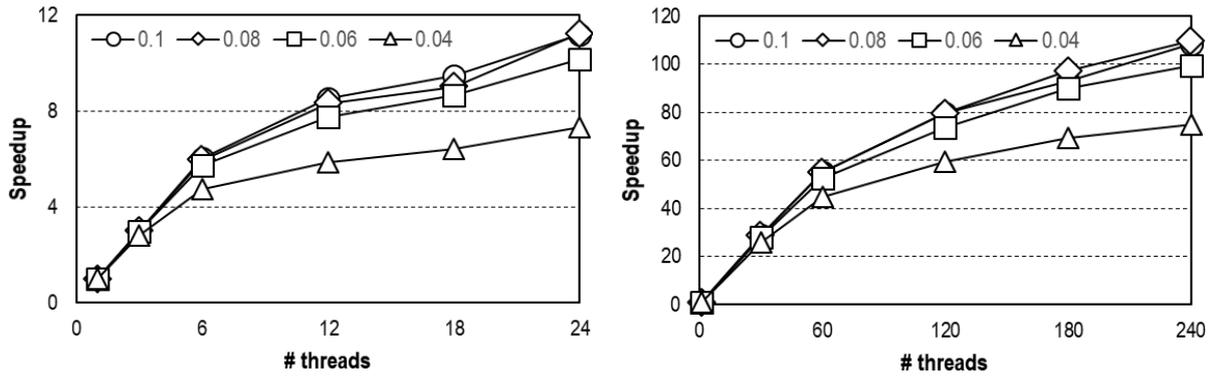

a) 2×Intel Xeon

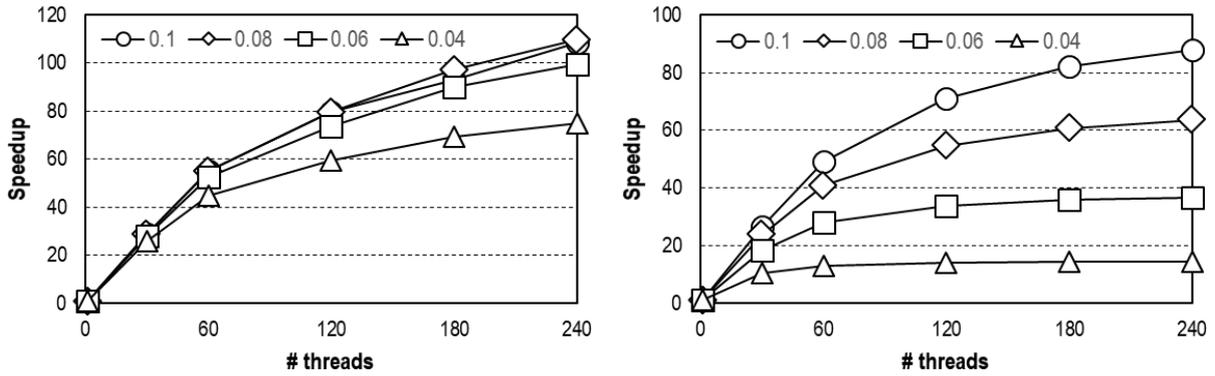

b) Intel Xeon Phi

*Figure 6.* Speedup of the algorithm on Tornado20M (left) and 20M (right) datasets w.r.t. the $minsup$ threshold.

5.3. Discussion

In this paper, we propose a parallel version of the DIC algorithm for Intel Xeon and Xeon Phi many-core systems and exploit a direct bit representation of both transactional database and itemsets. Our implementation codes a transaction or an itemset as a 64-bit integer, i.e. $m$, the number of items in the problem statement, is limited by 64. This limitation is clearly unacceptable for some applications, e.g. search for items that frequently purchased together by customers in a supermarket, search for frequent DNA sequences, and so on. However, the following brief review of papers shows that our algorithm is applicable for discovering interesting association rules in medical data. Li *et al.* in [13] proposed a method for mining optimal risk pattern sets and evaluated the algorithm on two real medical datasets with less than 30 attributes. In [14] and [15], Ordonez *et al.* introduced an algorithm to discover association rules in medical data, which incorporates several important constraints. Authors described how medical records were mapped to a transactional format suitable for mining. In the experiments, authors took at most 25 attributes of more than 100 patient's attributes since the chosen attributes provide a complete picture of the patient. In addition, the authors' experience showed that rules with more than 5 medical attributes were hard to interpret. At last, Pattanaprateep *et al.* in [17] described mining the association rules in the hospital database with more than 2.5 million records of patients' visits including attributes regarding patient's demographics, diagnose, and drug utilization.

## 6. Conclusion

In this paper, we have presented ParallelDIC, a parallel implementation of Dynamic Itemset Counting (DIC) algorithm for Intel many-core systems. DIC is a variation of classical Apriori algorithm for frequent itemset mining. We parallelize the DIC algorithm through OpenMP technology and thread-level parallelism. We propose the direct bit representation for transactions and itemsets with the assumption that such a representation of the transactional database fits in main memory. This technique reduces memory space for storing the transactional database, simplifies the support count via logical bitwise operation, and provides vectorization of this step. Our algorithm balances the support count between threads depending on the current total number of candidate itemsets. We performed an experimental evaluation on the platforms of the Intel Xeon CPU and the Intel Xeon Phi coprocessor with large synthetic and real databases (about millions of transactions), showing the good performance and scalability of the proposed algorithm, especially on the Intel Xeon Phi system.

However, it should be remembered that since ParallelDIC exploits the direct bit technique, this limits the number of items in the problem statement to 64. Nevertheless, literature review shows that despite this limitation, our algorithm is applicable for discovering interesting association rules in large medical datasets.

**Acknowledgments.** This work was financially supported by the Russian Foundation for Basic Research (grant No. 17-07-00463), by Act 211 Government of the Russian Federation (contract No. 02.A03.21.0011) and by the Ministry of education and science of Russian Federation (government order 2.7905.2017/8.9). Author thanks Lyudmila Kuznetsova and Valentin Komkov for their valuable comments on application of the algorithm in medicine.